\documentstyle[12pt]{article}

\csname @addtoreset\endcsname{equation}{section}

\def\bseq{\begin{subequation}}  % = 1a 1b
\def\eseq{\end{subequation}}
\def\bsea{\begin{subeqnarray}}  % = 1.1a 1.1b
\def\esea{\end{subeqnarray}}
\newcommand{\bbox}{\lower.2ex\hbox{$\Box$}}

\newcommand{\beq}{\begin{equation}}
\newcommand{\eeq}{\end{equation}}
\newcommand{\bea}{\begin{eqnarray}}
\newcommand{\eea}{\end{eqnarray}}
\newcommand{\ena}{\end{eqnarray}}

\renewcommand{\a}{\alpha}
\renewcommand{\b}{\beta}

\renewcommand{\d}{\delta}

\newcommand{\pa}{\partial}

\newcommand{\m}{\mu}

\newcommand{\n}{\nu}

\renewcommand{\P}{\Pi}

\newcommand{\s}{\sigma}
\renewcommand{\S}{\Sigma}

\newcommand{\Db}{\bar{D}}

\newcommand{\Sigmab}{\bar{\Sigma}}
\newcommand{\Phib}{\bar{\Phi}}

\newcommand{\ad}{{\dot{\alpha}}}
\newcommand{\bd}{{\dot{\beta}}}

\begin{document}

\begin{titlepage}
\begin{flushright} IFUM-603-FT \\ KUL-TF-98/03
\end{flushright}
\vfill
\begin{center}
{\LARGE\bf Duality in N=2 nonlinear sigma--models \footnote{To appear 
in the proceedings of {\em Quantum aspects of gauge theories, 
supersymmetry and unification}, Neuchatel University, 18--23 September 1997}}\\
\vskip 27.mm  \large
{\bf   Silvia Penati $^1$, Andrea Refolli $^1$,\\
Antoine Van Proeyen $^{2}$ and  Daniela Zanon $^1$ } \\
\vfill
{\small
 $^1$ Dipartimento di Fisica dell'Universit\`a di Milano and\\
INFN, Sezione di Milano, via Celoria 16,
I-20133 Milano, Italy\\
\vspace{6pt}
$^2$ Instituut voor theoretische fysica,
Katholieke Universiteit Leuven,\\ B-3001 Leuven, Belgium. }
\end{center}
\vfill

\begin{center}
{\bf ABSTRACT}
\end{center}
\begin{quote}
We consider $N=2$ supersymmetric nonlinear sigma--models in two
dimensions defined in terms of the nonminimal scalar multiplet.
We compute in superspace the one--loop beta function and show 
that the classical duality between these models and the standard ones
defined in terms of chiral superfields is maintained at the quantum
one--loop level. Our result provides an explicit application
of the recently proposed quantization of the nonminimal scalar
multiplet via the Batalin--Vilkovisky procedure.
\end{quote}
\end{titlepage}

\section{Introduction}
 
Since the seminal paper by Seiberg and Witten \cite{b1} who computed exactly
the low energy effective action for $N=2$ supersymmetric Yang--Mills 
theories by exploiting a duality symmetry
of the moduli space, duality has been playing a central role in the 
current research in quantum field theory and strings.
In a string theory contest two dimensional nonlinear sigma--models 
appear as classical string vacua. The study of dualities 
between different classes of nonlinear sigma--models may give  
insights on the relations between different geometries underlying 
string compactifications.

Here we consider $N=2$ nonlinear 
sigma--models defined in terms of complex linear superfields \cite{b2,b3}.
It is well known \cite{b3} that they are 
classically dual to the 
standard nonlinear sigma--models defined in terms of chiral superfields. 
The duality which relates these two sets of
models does not require the existence of a target space isometry, 
in contradistinction with the case of $N=2$ chiral,
twisted--chiral and semi--chiral sigma--models where an abelian isometry
is needed in order to realize a duality symmetry
(see Grisaru, Massar, Sevrin, Troost's contribution to these
proceedings for an exhaustive review and Ref. \cite{GMST}).
While at the classical level dual models represent a different 
parametrization of the same theory, the question is whether this property
survives after quantization as a property of the full 
quantum actions. In other words, the question is whether the duality 
transformations commute with the renormalization procedure. 
Since the renormalization properties of sigma--models are encoded in 
their beta functions,
we have computed the one--loop beta function for
the sigma model defined in terms of complex linear superfields and
compared it with the standard beta function for the chiral case \cite{b4,b5}.
As a result we have obtained that at the one--loop level, 
classical duality is not affected by renormalization.
The calculation has been performed by using the recently proposed \cite{b6}   
quantization of the complex linear multiplet.   
The linearity constraints are solved in terms of unconstrained gauge 
superfields and, consequently, an infinite number of gauge invariances 
appear. They can be consistently gauge--fixed by using the 
Batalin--Vilkovisky approach \cite{b7}. This procedure leads to a
gauge--fixed action which contains an infinite tower
of ghosts interacting among themselves and with the physical fields.
However we will show that, by choosing gauge--fixing functions 
independent of the physical background, the infinite tower of ghosts 
decouple in the process of diagonalizing the quadratic action. 
Eventually we obtain a simple expression for the propagators of the
fields that enter the perturbative calculations. 

The paper is organized as follows: in Section 2 we review some basic
notions concerning the chiral and complex linear superfields
and the classical duality 
between sigma--models defined in terms of chiral superfields and
complex linear ones. The quantization of a sigma model 
with complex linear superfields is presented in Section 3 where we
 explain how to deal with the infinite tower of ghosts appearing 
in the gauge--fixed action. Moreover the propagator for the physical fields
is obtained.
The one--loop beta function is computed in Section 4
following Ref. \cite{b10}. The comparison with the standard one--loop 
result for the chiral sigma--model is then discussed.
Finally, Section 5 contains our conclusions.
 
\section{Chiral and complex linear multiplets: the classical duality}
  
The standard description of $N=2$ supersymmetric scalar matter in two
dimensions is in 
terms of a minimal multiplet, the chiral multiplet, defined by the  
constraints $\bar{D}_{\dot{\a}} \Phi =0$, $D_{\a} \bar{\Phi} =0$.
The free action 
\begin{equation}
S = \int d^2 x d^2 \theta d^2\bar{\theta} ~\bar{\Phi} \Phi 
= \int d^2x \left[ \bar{\phi} \Box \phi - \bar{\psi}^{\dot{\a}} 
i\partial_{\a \dot{\a}} \psi^{\a} + \bar{F} F \right]
\label{1}
\end{equation}
describes the dynamics of a physical scalar $ \phi =\Phi|$, a spinor 
$ \psi_{\a} = D_{\a} \Phi|$ and an auxiliary field $F =D^2 \Phi|$. 
A different representation of $N=2$ supersymmetry is given by the
complex linear multiplet which satisfies the constraints 
$\bar{D}^2 \Sigma =0$, $D^2 \bar{\Sigma} =0$. 
It is a nonminimal reducible representation 
containing $12+ 12$ degrees of freedom \cite{b2,b3}. 
Since this multiplet can be
consistently coupled to Yang--Mills it can be used to describe
scalar matter. The free action
\begin{eqnarray}
S &=& \int d^2x d^2\theta d^2 \bar{\theta} ~\bar{\Sigma} \Sigma 
\nonumber \\
&=& \int d^2x \left[ \bar{B} \Box B - \bar{\zeta}^{\dot{a}} 
i\partial_{\a \dot{\a}} \zeta^{\a} -\bar{H} H + \beta^{\a} \rho_{\a}
+\bar{\beta}^{\dot{\a}} \bar{\rho}_{\dot{a}} + \bar{p}^{\a \dot{\a}}
p_{\a \dot{\a}} \right]
\label{2}
\end{eqnarray}
has the same physical content as the action (\ref{1}), 
$B=\S|$, $\zeta_{\a} = D_{\a} \bar{\S}|$, but a different  
auxiliary sector $\rho_{\a} = D_{\a} \S|$, $H = D^2 \S|$, $p_{\a \dot{\a}}
=\bar{D}_{\dot{\a}} D_{\a} \S|$ and $\bar{\b}_{\dot{\a}}=\frac12 D^{\a}
\bar{D}_{\dot{\a}} D_{\a} \S|$. 

The two actions are classically dual. Writing the first order 
action
\begin{equation}
S = \int d^2x d^2\theta d^2 {\bar{\theta}} \left[ \bar{\Phi} \Phi
+ \Sigma \Phi + \bar{\Sigma} \bar{\Phi} \right] 
\end{equation}
where $\Sigma$ is complex linear and $\Phi$ unconstrained, it is easy
to see that performing the gaussian integral in $\Phi$ we recover the
action (\ref{2}), whereas functional integration on $\sigma^{\a}$ and 
$\bar{\sigma}^{\dot{\a}}$ ($\Sigma = \bar{D}_{\dot{\a}} 
\bar{\sigma}^{\dot{\a}}$,
$\bar{\Sigma} = D_{\a} \sigma^{\a}$) leads to the action (\ref{1}).
We notice that under duality the chirality constraints $\bar{D}_\ad \Phi=0$
and the equations of motion $D^2 \Phi =0$ from the action (\ref{1}) 
are interchanged with the linearity constraints $D^2 \bar{\Sigma} =0$ 
and the equations of motion $\bar{D}_\ad \bar{\Sigma} =0$ from the action
(\ref{2}). This is similar to what happens in the standard electromagnetic
duality. 
Indeed this is not an accident since in four dimensions it has
been shown \cite{b9} that the standard $N=2$ supersymmetric Yang--Mills  
theory with chiral matter coupled to an abelian gauge sector is dual 
to an $N=2$ abelian Yang--Mills system where the scalar sector is 
described in terms of a complex linear multiplet. 
In this contest the $\Phi \rightarrow \bar{\S}$ duality
in the matter sector is accompanied by the electromagnetic duality in the
gauge sector. 

More generally we consider $N=2$ two dimensional nonlinear sigma--models 
in terms of chiral or complex linear superfields.
Given a set of superfields
$\Phi^\m $, $\Phib^{\bar{\m}}$, $\S^\m $, $\Sigmab^{\bar{\m}}$ with
$\m,\bar{\m}=1,\dots,n$ we write the first order action
\begin{equation}
S= \int d^4x~ d^4\theta~ \left[ K(\Phib, \Phi) + \S \Phi +\Sigmab \Phib
\right]
\label{actionsigma}
\end{equation}
where $\S$ are linear superfields, whereas $\Phi$ are initially
unconstrained (we simplify the notations by obmitting 
the indices $\m$, $\bar{\m}$ on the superfields).
Performing the functional integral on $\S$ we obtain 
the chirality constraint on $\Phi$, so that the quadratic terms, 
being total derivatives, can be dropped and one is left with the 
standard sigma--model action for chiral superfields with K\"ahler 
potential $K$.
The dual model is defined as a Legendre transform of the previous   
theory. One solves the classical equations of  
motion for $\Phi$ 
\begin{equation}
\S=-\frac{\pa K}{\pa \Phi}\qquad \qquad \Sigmab=-\frac{\pa K}{\pa \Phib}
\end{equation}
obtaining $\Phi=\Phi(\S,\Sigmab)$, $\Phib=\Phib(\S,\Sigmab)$. The 
dual action is then given by the action (\ref{actionsigma}) once 
these solutions are inserted
\begin{equation}
S =\int d^4x~ d^4\theta~ \tilde{K}(\S,\Sigmab)
\label{dualaction}
\end{equation}
where 
\begin{equation}
\tilde{K}(\S,\Sigmab)= [K(\Phi,\Phib)+\S\Phi+\Sigmab
\Phib]|_{\Phi=\Phi(\S,\Sigmab), \Phib=\Phib(\S,\Sigmab)}
\label{dualpot}
\end{equation}
The classical duality of these theories is well understood in a 
superspace formulation where all the fields are off-shell. Since the
two multiplets have different auxiliary fields, going to components 
and eliminating the auxiliary fields through their equations of
motion might lead to quite different results in the two cases \cite{deo}. 
Thus, when investigating the duality properties at the 
quantum level, it is convenient to proceed in a completely off--shell,
superspace formulation . 

Going from one model to its dual amounts to replace the potential
$K$ with its Legendre transform $\tilde{K}$ in eq. (\ref{dualpot}). 
If we define the second derivatives matrix for $K$
\begin{equation}
G=\left( \matrix{\frac{\pa^2 K}{\raise-.04cm\hbox{$\scriptstyle
\pa \Phi \pa \Phib$}}&
\frac{\pa^2 K}{\raise-.04cm\hbox{$\scriptstyle
\pa \Phi \pa \Phi$} } \cr
~~&~~ \cr
\frac{\pa^2 K}{\raise-.04cm\hbox{$\scriptstyle
\pa \Phib \pa \Phib$}} &
\frac{\pa^2 K}{\raise-.04cm\hbox{$\scriptstyle
\pa \Phib \pa \Phi$}} \cr}
\right)
\label{metric}
\end{equation}
and a similar matrix for $\tilde{K}$, it is easy to show that under
duality transformation
\begin{equation}
G(\Phi,\Phib)\rightarrow   -\tilde{G}(\S,\Sigmab) = G(\Phi,\Phib)^{-1}
\label{dualmetric}
\end{equation}
In order to investigate the issue of quantum duality we will study 
whether the relation (\ref{dualmetric}) is maintained at the one--loop
order.
 
\section{The quantum approach}

Since the duality transformation 
corresponds to performing functional integrals over the fields in two
different orders, in principle it could be affected by the
quantization procedure. In a perturbative approach the obvious  
question to ask is whether the duality properties are maintained order by order 
under renormalization. This can be established by studying the relation 
(\ref{dualmetric}) beyond tree level. The simplest 
object which can be computed perturbatively and is a function of the
matrix elements in (\ref{metric}) is the $\b$-function. 
In the case of a standard $N=2$ sigma--model the one--loop divergent
contribution to the K\"ahler potential and the corresponding 
$\b$--function are given respectively by the well known results
\begin{equation}
K^{(1)} = \frac{1}{\epsilon} {\rm{tr}} \log K_{\m \bar{\n}}
\label{1loopdiv}
\end{equation}
\begin{equation}
\b_{\m \bar{\m}} = -\pa_{\m} \pa_{\bar{\m}} {\rm {tr}} \log{K_{\n \bar{\n}}}
= -R_{\m \bar{\m}}
\label{beta}
\end{equation}
We concentrate on the calculation of the first order correction to 
the potential $\tilde{K}$ in the case of a sigma--model defined in
terms of complex linear superfields. Following the standard procedure
for the chiral case \cite{b5} we perform the calculation by using a
quantum--background splitting $\S \rightarrow \S_0 + \S$, 
$\bar{\S} \rightarrow \bar{\S}_0 + \bar{\S}$ and compute one--loop
diagrams with external background lines with no derivatives acting
on them. From the expansion of the action (\ref{dualaction}) around
the classical background
\bea
S &=& \int d^2x d^4\theta \left\{ -\S^{\m} \bar{\S}^{\bar{\n}} 
\delta_{\m \bar{\n}} + [\tilde{K}_{\m \bar{\n}}(\S_0,\bar{\S}_0) +
\delta_{\m \bar{\n}}] \S^{\m} \bar{\S}^{\bar{\n}} +\frac12
\tilde{K}_{\m \n}(\S_0,\bar{\S}_0) \S^{\m} \S^{\n} \right. \nonumber \\ 
&~&~~~~~~~~~~~~~~ 
\left. +\frac12 \tilde{K}_{\bar{\m} \bar{\n}}(\S_0,\bar{\S}_0) 
\bar{\S}^{\bar{\m}} \bar{\S}^{\bar{\n}} + \cdots \right\} 
\label{split}
\ena
we can read the vertices to be used in one--loop Feynman diagrams.
However, in order to perform the calculation we need to quantize 
consistently the constrained $\S$ and $\bar{\S}$ superfields. 

We use the recently proposed \cite{b6} quantization of the complex 
linear multiplet expressed in terms of two unconstrained 
spinor prepotentials
\begin{equation}
\S = \bar{D}_{\dot{\a}} \bar{\s}^{\dot{\a}} \qquad \qquad 
\bar{\S} = D_{\a} \s^{\a}
\label{prepot}
\end{equation}
The quadratic action 
\begin{equation}
S = -\int d^2x d^4 \theta \bar{\s}^{\dot{\a}} \bar{D}_{\dot{\a}} 
D_{\a} \s^{\a}
\label{starting}
\end{equation}
is infinitely reducible and the
Batalin--Vilkovisky gauge--fixing procedure introduces an infinite
number of ghosts. We choose gauge--fixing functions independent 
of the background so that the ghosts couple to the quantum fields $\s^{\a}$ and 
$\bar{\s}^{\dot{\a}}$, but not
to the physical background. Starting from the action
(\ref{starting}) and performing the gauge--fixing as explained
in \cite{b6,b10} one obtains a final quadratic gauge--fixed action with
all the diagonal terms invertible and an infinite number of nondiagonal
terms mixing physical fields and ghosts. In order to perform
perturbative calculations we need to compute the
propagators for the prepotentials in (\ref{prepot}) and these can be 
read from the gauge--fixed action once the diagonalization is 
performed. Since we are dealing with an infinite number of terms
this looks like a very hard task . However in \cite{b10} it has
been shown that a partial diagonalization is sufficient to 
disentangle the physical fields $\s^{\a}$ and $\bar{\s}^{\dot{\a}}$
from the rest of the ghost action. Therefore, the ghosts, still
interacting among themselves, in a highly nontrivial way, completely decouple 
from the physical sector. 
This procedure modifies the diagonal part of the 
action for $\s^{\a}$ and $\bar{\s}^{\dot{\a}}$. 
From its explicit expression (see Ref. \cite{b10}) one 
obtains the following propagator
\begin{eqnarray}
<\s^\a \bar{\s}^\ad>&=&
{(\tilde{W}^{-1})}^{\a \dot \a}= - \frac{i \pa^{\a \dot \a}}{\Box}
+\frac{3(kk'_1)^2 + 4 -2k^{'2}_1}{4(kk'_1)^2}
i \pa^{\a \dot \a} \frac{D^2 \bar D^2}{\Box^2}+
\nonumber \\
&&+
\frac{3 k^2 -2}{4 k^2} i \pa^{\a \dot \a} \frac{D_\b \bar D^2 D^\b}{\Box^2}+
\frac{2-k^2}{4 k^2} i \pa^{\a \bd} i \pa^{\b \ad}
\frac{D_\b  \Db_\bd}{\Box^2}
\label{propag}
\end{eqnarray}
where $k$ and $k_1'$ are gauge parameters introduced in the gauge--fixing
procedure.

\section{One--loop beta function}

Now we have collected all the ingredients which are necessary
for the computation of the one-loop $\b$-function for
the linear multiplet sigma--model.  
As shown in the action (\ref{split}), the superfields 
$\s^{\a}$ and $\bar{\s}^{\dot{\a}}$ always
couple to the external background through their field strengths
$\S$, $\bar{\S}$. Therefore, only the $<\bar{\S} \S>$ propagator 
enters in the perturbative calculations. 
From the propagator in (\ref{propag}) we obtain
\beq
<\Sigmab\S>=D_\a<\s^\a \bar{\s}^\ad>\Db_\ad =
  \frac{ D^2 \bar  D^2}{\Box}+
\frac{D_{ \a}  \Db^2  D^{\a}}{\Box}
\equiv \P
\label{risfin}
\eeq
This expression is independent of any gauge parameter 
introduced in the gauge-fixing procedure and this provides
a consistency check of the methods used in the quantization of
the linear multiplet.
From the action (\ref{split}) we define
\beq
{\cal V}\equiv (\tilde{K}_{\m\bar{\n}}+ \d_{\m\bar{\n}})\qquad,
\qquad{\cal U}\equiv \tilde{K}_{\m\n}
\label{def}
\eeq
It is easy to see that 
one--loop divergent contributions to the effective action for the
linear multiplet sigma--model are of two types:
the ones corresponding to graphs which contain only $\S {\cal V} \bar{\S}$
interactions and the ones associated to diagrams with both $\S {\cal V} \bar{\S}$  
and $\S {\cal U} \bar{\S}$, $\bar{\S}\bar{{\cal U}}\bar{\S}$ vertices.  

In order to compute the divergent contributions we use standard
$D$-algebra techniques very similar to the ones used in \cite{b5}.
The details of the calculations can be found in Ref. \cite{b10}.
Here we give only the final results. 
From the first set of diagrams one obtains
\beq
\tilde{K}^{(1)}_1\rightarrow 
-\frac{1}{\epsilon} \rm{tr} \log(1-{\cal V})
\label{F1}
\eeq
while the sum of the infinite set of one--loop diagrams containing 
any number of ${\cal V}$ vertices and an equal number of ${\cal U}$
and $\bar{\cal U}$ vertices gives 
\beq
\tilde{K}^{(1)}_2\rightarrow -\frac{1}{\epsilon} \rm{tr}
\log(1-{\cal U} \frac{1}{1-{\cal V}}
\bar{{\cal U}}\frac{1}{1-{\cal V}})
\label{F2}
\eeq
Adding the results in (\ref{F1}) and (\ref{F2}) we obtain the
total one-loop divergent contribution
\beq
\tilde{K}^{(1)} = -\frac{1}{\epsilon} \left[
\rm{tr} \log(1-{\cal V})+\rm{tr}
\log(1-{\cal U} \frac{1}{1-{\cal V}}
\bar{{\cal U}}\frac{1}{1-{\cal V}}) \right]
\label{Ftotal}
\eeq
or, in terms of the second derivatives of the potential, 
\beq
\tilde{K}^{(1)} = -\frac{1}{\epsilon}
\rm{tr} \log[-(\tilde{K}_{\m\bar{\m}}-\tilde{K}_{\m\n}
\tilde{K}^{-1}_{\n\bar{\n}}
\tilde{K}_{\bar{\n}\bar{\m}})]
\eeq
Using the definition (\ref{metric}) for the second derivative matrix of 
$\tilde{K}$, it can be rewritten as
\beq
\tilde{K}^{(1)} = \frac{1}{\epsilon} {\rm{tr}} \log (-\tilde{G})_{\m \bar{\n}}
\label{FFtotal}
\eeq
Now we compare the above expression with the result in (\ref{1loopdiv}), 
where the one-loop divergent
contribution to the K\"ahler potential of the $N=2$ theory is exhibited.
On the basis of the classical correspondence in (\ref{dualmetric}),
we conclude that the results in (\ref{1loopdiv})
and (\ref{FFtotal}) explicitly maintain the expected duality.
 
\section{Conclusions}

We have computed the one--loop divergent contribution to the potential
for a nonlinear sigma--model defined in terms of complex linear superfields
and proved that classical duality with the standard chiral sigma--model
is maintained at one--loop order. 

We performed the calculation in the quantization scheme recently proposed
in \cite{b6}. We have shown that, even if in general that 
method might be difficult to use in applications due to the 
presence of an infinite chain of ghosts, in our case an intelligent choice of the
gauge--fixing terms allows for the complete decoupling of the ghosts from
the physical sector as in standard abelian gauge theories. Further 
applications in quantum 
supersymmetric theories containing complex linear superfields (see for 
instance \cite{b9}) might be viable. 

Quantum duality at higher loop levels needs to be studied:
there duality transformations could
receive nontrivial quantum corrections as in the more familiar case of
dual bosonic sigma--models \cite{b11}.
Moreover, our result can be easily extended \cite{b12}
to mixed nonlinear sigma--models
defined in terms of both chiral and complex linear superfields which 
naturally arise in supersymmetric extensions of QCD low--energy effective 
action \cite{b13,b14}.

The geometry underlying a standard $N=2$ chiral sigma--model is
K\"ahler, while a complete geometric interpretation is still lacking
for the dual theory (however see ref. \cite{b14}). A detailed discussion 
of this issue can be found in \cite{b12}.  

\vskip 15pt
\noindent   
{\bf Acknowledgments}

\noindent
This work was supported in part by MURST and by the European Commission 
TMR programme 
ERBFMRX--CT96--0045, in which S.P, A.R. and D.Z. are associated to
the University of Torino. A.V.P. thanks the FWO Belgium for financial
support.  
%%%%%%%%%%%%%%%%%%%%%%

\end{document}